\documentclass[fleqn,usenatbib]{mnras}
\usepackage{array}
\usepackage{newtxtext,newtxmath}
\usepackage[T1]{fontenc}

\usepackage[flushleft]{threeparttable}
\usepackage[mathlines,switch]{lineno}
\usepackage{kotex}
\usepackage{placeins}
\usepackage{diagbox}
\usepackage{makecell}
\usepackage{tabularx}
\usepackage{bm}
\usepackage{graphicx}
\usepackage{float}
\newcolumntype{Y}{>{\centering\arraybackslash}X}

\usepackage[usenames,dvipsnames]{xcolor}

\usepackage{soul}

\title[Enhancing the Detection Sensitivity of Primordial PV]{Enhancing the Detection Sensitivity of Primordial Parity Violation using Galaxy Spins}

\author[B. J. Yu et al.]{
Byoungjo Yu$^{1}$,
Junsup Shim$^{2,3}$\thanks{E-mail: jsshim@pusan.ac.kr},
and Hyunmi Song$^{1}$\thanks{E-mail: hmsong@cnu.ac.kr}
\\
$^{1}$Department of Astronomy and Space Science, Chungnam National University, Daejeon 34134, Republic of Korea\\
$^{2}$Department of Earth Sciences, Pusan National University, Busan 46241, Republic of Korea\\
$^{3}$Academia Sinica Institute of Astronomy and Astrophysics (ASIAA), No. 1, Section 4, Roosevelt Road, Taipei 10617, Taiwan
}

\begin{document}

\maketitle

\begin{abstract}
It has been recently demonstrated that the signature of primordial parity violation could be imprinted in halo spins, indicating its potential detectability through the late-time galaxy spin field (Shim et al. 2025). In this study, we develop an optimized halo selection strategy to enhance the detection significance of such a signal, focusing on halo mass and local density. Using $N$-body simulations with parity-asymmetric initial conditions, we show that the optimized halo sample allows for a higher detection sensitivity than the full halo sample, despite its reduced sample size. Finally, we assess the observational feasibility of our strategy and show that future spectroscopic surveys can provide sufficient data to realize this enhanced sensitivity.
\end{abstract}

\section{Introduction}\label{sec:intro}
The symmetry properties of cosmic initial conditions, along with their statistical properties, offer fundamental insights into the early universe physics. Traditionally, the statistical properties of initial conditions have been constrained using two-point information from scalar observables \citep{Bardeen_1986, Bond_1987, Eisenstein_1999, Bridle_2003, Desjacques_2008, Matarre_2008, Hinshaw_2013, Vieira_2018, Shim_2021, Feldbrugge_2023, Chudaykin_2026}, such as the fluctuation fields of the cosmic microwave background temperature and galaxy (or matter) density, supplemented by their higher-point statistics \citep{Maldacena_2003, Eiichiro_2001, Dalal_2008, Bennett_2013, Koyama_2018, Akrami_2020, Cabass_2022}. However, testing parity symmetry with scalar fields is inherently challenging, because scalar quantities lack intrinsic `\textit{handedness}'; the simplest parity-odd observable becomes a four-point correlation function to define a non-zero helical volume in three dimensions \citep{Philcox_2022, Hou_2023, Philcox_2023, Greco_2025}.

Recent measurements of the galaxy four-point correlation function \citep{Philcox_2022, Hou_2023} have sparked growing interest in searching for signatures of broken parity symmetry in the large-scale structure (LSS) of the Universe. Despite this progress, the high-order nature of these statistics complicates the interpretation of results \citep{Hou_2025, Philcox_2024, Slepian_2025}. Specifically, the covariance analysis for four-point functions is substantially more complex than for two-point statistics, making it difficult to isolate primordial signals from systematic or late-time effects. 
Consequently, a multi-faceted approach leveraging non-scalar observables is essential to establish and interpret any potential signal of parity asymmetry in the LSS.\footnote{Complementary efforts have also focused on compressing four-point information into more manageable lower-point statistics \citep{Jamieson_2024, Gao_2026}.}

A promising alternative to scalar observables involves higher-spin quantities such as vector and tensor observables of LSS, including galaxy spin vectors (spin-1) \citep{Yu_2020, Motloch_2022, Coulton_2024, Shim_2025}, and galaxy shape alignment and tidal force tensors (spin-2) \citep{Philcox_2024b, Okumura_24, Kurita_2025, Mikura_2025, Zhu_2025}. The intrinsic handedness of these higher-spin fields allows their parity-violating signatures to be captured at the two-point level. This significantly simplifies the associated covariance analysis, while providing independent and complementary constraints. These probes act as crucial independent cross-checks to scalar measurements, forming a coherent framework for testing the fundamental symmetries of the Universe.

In this context, \cite{Shim_2025} recently demonstrated that primordial parity violation leaves a detectable imprint on the late-time angular momentum directions of galaxies -- the galaxy spin vector field -- accessible through its power spectrum. In such scenarios, the initial tidal torque field — arising from the misalignment between the initial tidal and inertia tensors, responsible for the generation of halo angular momentum via tidal torquing \citep{Doroshkevich_1970, White_1984} — becomes parity-asymmetric. This primordial asymmetry is inherited by the spin fields of protohalos and persists into the late-time halo population. However, nonlinear gravitational evolution tends to decorrelate the late-time spin field from its initial state, weakening the observable signal. To maximize the discovery potential of this probe, it is therefore desirable to identify halo populations that best preserve the primordial symmetry-breaking signature.

In this paper, we develop a method to enhance the detection significance of parity violation using galaxy spins through the construction of an optimal halo sample. Unlike the previous analysis that utilized the full halo sample, we focus on identifying subsets of halos that better preserve their primordial spin information and are therefore more sensitive to parity-violating signatures. We investigate halo mass and local density as the key environmental parameters that determine the degree of alignment between the initial and late-time spin, thereby guiding the selection of an optimal galaxy sample for symmetry tests.

The paper is organized as follows. Section~\ref{method} describes the simulation data, the construction of the halo spin field, and the measurement of its helical asymmetry.
In Section~\ref{Mass dependence of primordial parity-violating signals}, we first examine how the detectability of the parity-violating signal changes with halo mass thresholds, while Section~\ref{sec: Signal enhancement through early-late halo spin alignment selection} extends this selection criterion by exploring the alignment between early- and late-time spins, focusing on halo mass and local density as observable proxies to amplify the signal beyond conventional mass-cut approaches.
We discuss in Section~\ref{sec: Observation of the PV Signal} the observational feasibility of these strategies, showing that current and future spectroscopic surveys can provide the data necessary to realize this enhanced sensitivity. Finally, we conclude in Section~\ref{sec:summary}.
Throughout this paper, we use a flat $\Lambda$CDM cosmology with $H_0=72\,{\rm km\,s^{-1}\,Mpc^{-1}}$ and $\Omega_m=0.258$.

\section{Method}\label{method}
\subsection{Simulation}
We utilize dark-matter-only $N$-body simulations used in \citet{Shim_2025}, where a total of $256^3$ dark matter particles were simulated within a periodic volume of  $100^3\,h^{-3}\,{\rm Mpc}^3$. There are two types of simulations, with 500 realizations of each type: one group with parity-symmetric initial conditions and the other with parity-asymmetric initial conditions. Hereafter, we refer to these as the parity-symmetric (PS) and parity-asymmetric (PA) simulations, respectively.
The standard PS initial conditions are set using the Zel'dovich displacement \citep{Zeldovich_1970}, while PA cases adopt a modified displacement that involves purely right-handed helical displacement in addition to the Zel'dovich displacement. The two suites of simulations share the identical initial matter power spectrum. For additional details of the simulations, we refer the reader to \citet{Shim_2025}.

\subsection{Methodology}\label{methodology}
We construct a halo spin field, $\mathbf{J}(\mathbf{x})$, at redshift $z=0$ by placing the unit spin vector of each halo onto a regular grid of $\sim 0.39\,h^{-1}\,\mathrm{Mpc}$, reflecting the challenge of measuring the actual magnitude of the spin. The unit spin vector of a halo is computed from its member particles as
\begin{equation} \label{eq:halo_spin}
\mathbf{j}_{\text{halo}} \equiv \frac{\sum_{i} \mathbf{r}^{i}_{p} \times \mathbf{v}^{i}_{p}}{\left| \sum_{i} \mathbf{r}^{i}_{p} \times \mathbf{v}^{i}_{p} \right|}\,,
\end{equation}
where $\mathbf{r}^{i}_{p}$ and $\mathbf{v}^{i}_{p}$ are the position and velocity of the $i$-th particle relative to the halo center.

We extract two opposite helical modes of the halo spin field to measure its helical asymmetry. In Fourier space, its right(R)-/left(L)-helical modes are computed through
\begin{equation}
    \mathbf{J}_{R/L}(\mathbf{k})=\mathbb{P}_{R/L}(\mathbf{k})\mathbf{J}(\mathbf{k}),
\end{equation}
where the projection matrix is constructed as 
\begin{equation}
\mathbb{P}_{R/L}=\hat{e}_{R/L}(\mathbf{k})\hat{e}_{R/L}^\dagger(\mathbf{k}).
\end{equation}
Here, $\hat{e}_{R/L}$ is the eigenvector of the curl operator in Fourier space given as,

\begin{equation}
\hat{e}_{R/L}(\mathbf{k}) = \frac{1}{\sqrt{2}k\sqrt{k_y^2 + k_z^2}} \left( \begin{array}{c} k_y^2 + k_z^2 \\ - k_x k_y \pm i k k_z \\ - k_x k_z \mp i k k_y \end{array} \right).
\end{equation}

The imbalance between the auto-spectra of the right- and left-helical spin fields, $P_{RR}(k)$ and $P_{LL}(k)$, is therefore a direct indicator of PV. We quantify the helical asymmetry as
\begin{equation} \label{eq:helical asymmetry}
\chi(k) \equiv
\frac{P_{RR}(k) - P_{LL}(k)}
     {P_{RR}(k) + P_{LL}(k)},
\end{equation}
with $\chi = 0$ for a parity-symmetric field and $|\chi| = 1$ for a maximally chiral one. We rescale $\chi(k)$ to mitigate the suppression of the asymmetry signal caused by residual shot noise in the denominator as in \citet{Shim_2025}. This adjustment compensates for the sensitivity of the denominator to field sparsity, ensuring that the rescaled measure reflects the intrinsic signal amplitude observed in the high-density limit.

\section{Mass dependence of primordial parity-violating signals} \label{Mass dependence of primordial parity-violating signals}
While selecting more massive halos is expected to better preserve primordial spin information \citep{Yu_2020}, it also reduces the available sample size, which can in turn degrade the statistical significance of the signal. To assess this trade-off and to identify an optimal mass threshold, we systematically investigate the dependence of the parity-violating signal on the halo mass cut. Specifically, we consider mass thresholds of $[10^{11}, 10^{11.5}, 10^{12}, 10^{12.5}]\, h^{-1}\,M_{\odot}$.

\begin{figure}
\includegraphics[width=\linewidth]{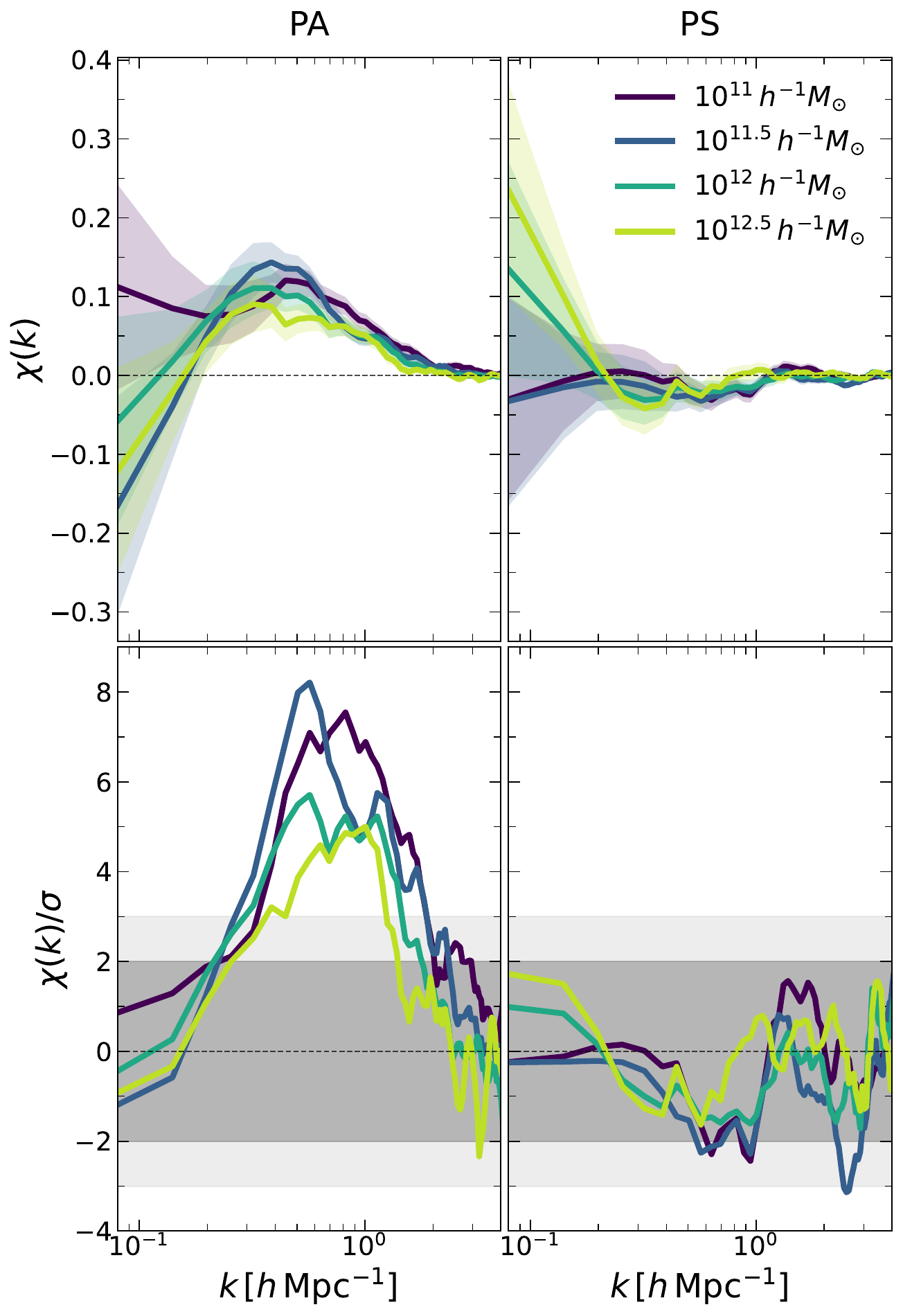}
\caption{Helical asymmetry power spectra (top) and their detection significance (bottom) for parity-asymmetric (PA, left) and parity-symmetric (PS, right) simulations. Different colors denote the results for various halo mass thresholds. Shaded regions in the top panels represent the standard error of the mean, while vertical lines mark the wavenumber corresponding to twice the Gaussian radius of each mass threshold. The gray horizontal bands in the bottom panels indicate the $<2\sigma$ and $<3\sigma$ significance ranges.}
\label{fig:masscut}
\end{figure}

In Figure~\ref{fig:masscut}, we show how the late-time parity asymmetry (top) and its detection significance (bottom) depend on the threshold mass of a halo sample in the PA (left) and PS (right) simulations. While the helical asymmetry remains comparable among different halo mass thresholds in the PA simulations, their detection significance shows a clear dependence on the halo sample.
The detection significance tends to increase toward lower mass thresholds, achieving the highest significance for $M\ge10^{11.5}h^{-1}M_{\odot}$, before slightly decreasing for the lowest mass threshold. This trend can be understood as the consequence of two competing effects toward lower mass thresholds: growing statistical gain from larger sample size versus the increasing number of halos with reduced primordial spin information (as we will show in Section~\ref{LE vs. significance}). Therefore, the highest detection significance is achieved when the two effects balance. Although the case for $M\ge10^{11.5}h^{-1}M_{\odot}$ yields the maximum significance, we adopt $M\ge10^{11}h^{-1}M_{\odot}$ as our fiducial threshold in the remainder of our analysis. This is because it provides the largest overall enhancement when our halo selection criterion is taken into account (see Section~\ref{sec: Signal enhancement through early-late halo spin alignment selection} and Figure~\ref{fig:LEallmass}).

\section{Signal enhancement through halo selection based on early-late spin alignment}\label{sec: Signal enhancement through early-late halo spin alignment selection}
Halos whose spin orientations are well preserved from early to late times provide a particularly sensitive probe of primordial parity-violating signals, because their spins retain the memory of the initial tidal torque field in which the primordial asymmetry, if present, is imprinted. In this section, we examine the properties of such halos and establish a practical criterion for distinguishing them that can be applied to observational data.

\subsection{Dependence of detection significance on early–late spin alignment} \label{LE vs. significance}
To quantify the change in halo spin direction between the early and late universe, we define the Lagrangian-Eulerian spin alignment (hereafter, LE alignment) as follows:
\begin{equation} \label{LE alignment}
\cos\theta_{\rm LE} = \hat{\mathbf{j}}_{\rm L} \cdot \hat{\mathbf{j}}_{\rm E},
\end{equation}
where $\hat{\mathbf{j}}_{{\rm L}}$ and $\hat{\mathbf{j}}_{{\rm E}}$ denote the halo spin direction measured in Lagrangian and Eulerian space, corresponding to the initial and late ($z=0$) spin, respectively. $\hat{\mathbf{j}}_{{\rm E}}$ and $\hat{\mathbf{j}}_{{\rm L}}$ are calculated directly from the member particle distribution using Eq.~\ref{eq:halo_spin}.

\begin{figure}
\includegraphics[width=\linewidth]{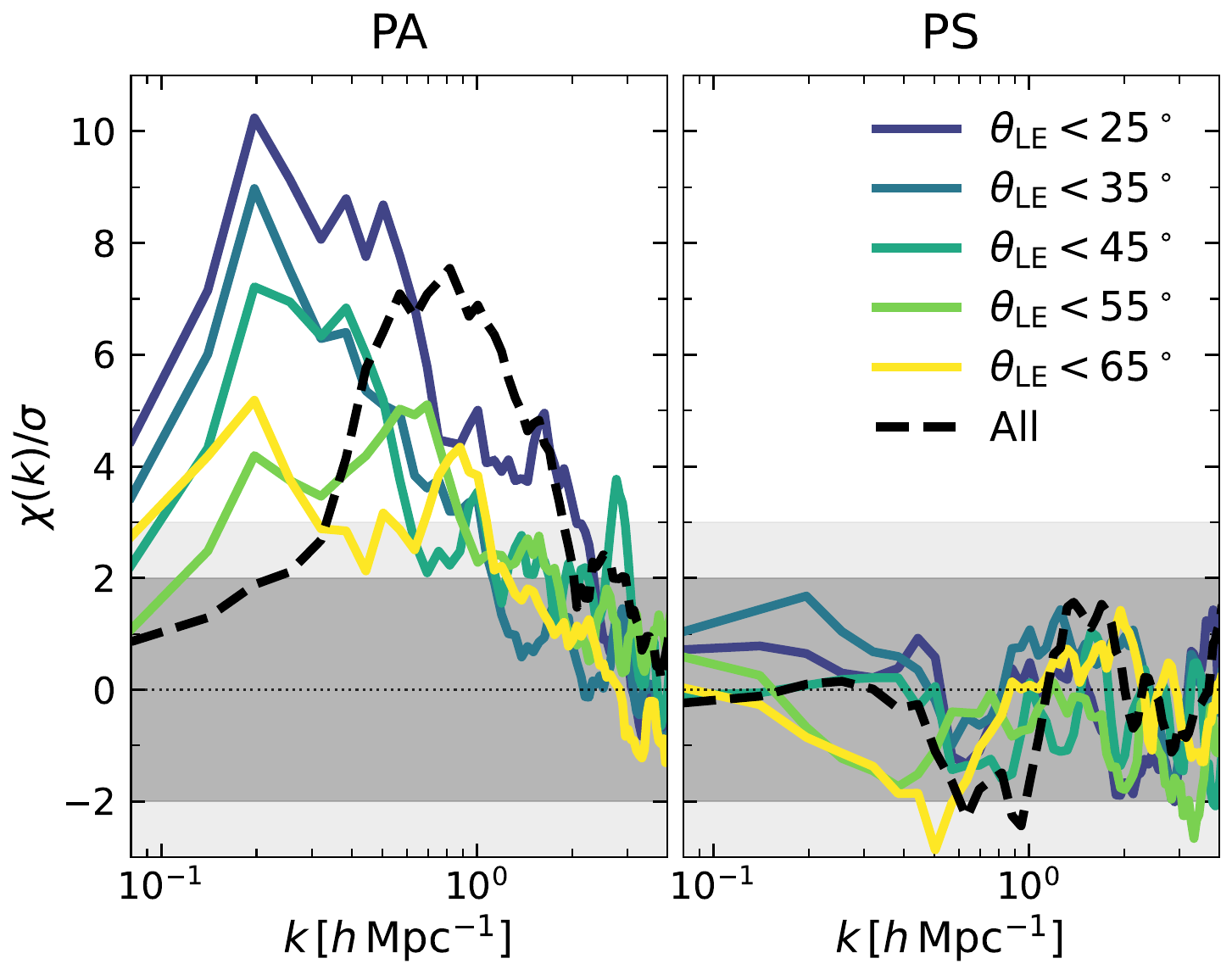}
\caption{Detection significance of the helical asymmetry for the full sample ($M_{\rm halo} \ge 10^{11}\,h^{-1}M_\odot$, black) and for subsamples with varying LE alignment thresholds. Each subsample size is matched to that of the smallest subsample (i.e., $\theta_{\rm LE}<25^\circ$) via random sampling. The left and right panels show results from the parity-asymmetric and parity-symmetric simulations, respectively.}
\label{fig:norm LE results}
\end{figure}

We investigate how the recovery of primordial parity information depends on the LE alignment in Figure~\ref{fig:norm LE results}. Halos of mass $\ge 10^{11}\, h^{-1}M_{\odot}$ are subsampled by varying the maximum allowed $\theta_{\rm LE}$ from 25$^\circ$ to 65$^\circ$ in steps of 10$^\circ$, and the helical asymmetry of the spin field (Eq.~\ref{eq:helical asymmetry}) is measured for each subsample. To isolate the effect of LE alignment from that of sample size, we construct the spin fields for all LE alignment thresholds using random subsamples matched in size to the halo subsample defined by the smallest maximum angle (25$^\circ$).

Halos with LE alignment angles smaller than $25^\circ$, representing only $\sim 16\%$ of the population, yield a signal approximately $2.7\sigma$ stronger than the full mass-limited sample in the PA simulations (left panel of Figure~\ref{fig:norm LE results}), while consistently remaining around $\chi(k)/\sigma\approx0$ in the PS cases (right panel of Figure~\ref{fig:norm LE results}).
This finding provides evidence that halos with stronger LE alignment serve as more sensitive tracers, effectively preserving primordial information against nonlinear evolution.
When considering relatively less aligned cases with $\theta_{\rm LE} < 45^\circ$, the significance becomes comparable to that of the full halo sample, and beyond this angle, the signal strength decreases further. This suggests that a LE alignment threshold of $\sim45^\circ$ serves as a practical transition threshold separating halos that largely preserve their primordial spin orientation from those that have experienced significant spin reorientation, and thus provides an effective criterion for maximizing sensitivity to the primordial parity-violating signal.

\begin{figure}
\includegraphics[width=\linewidth]{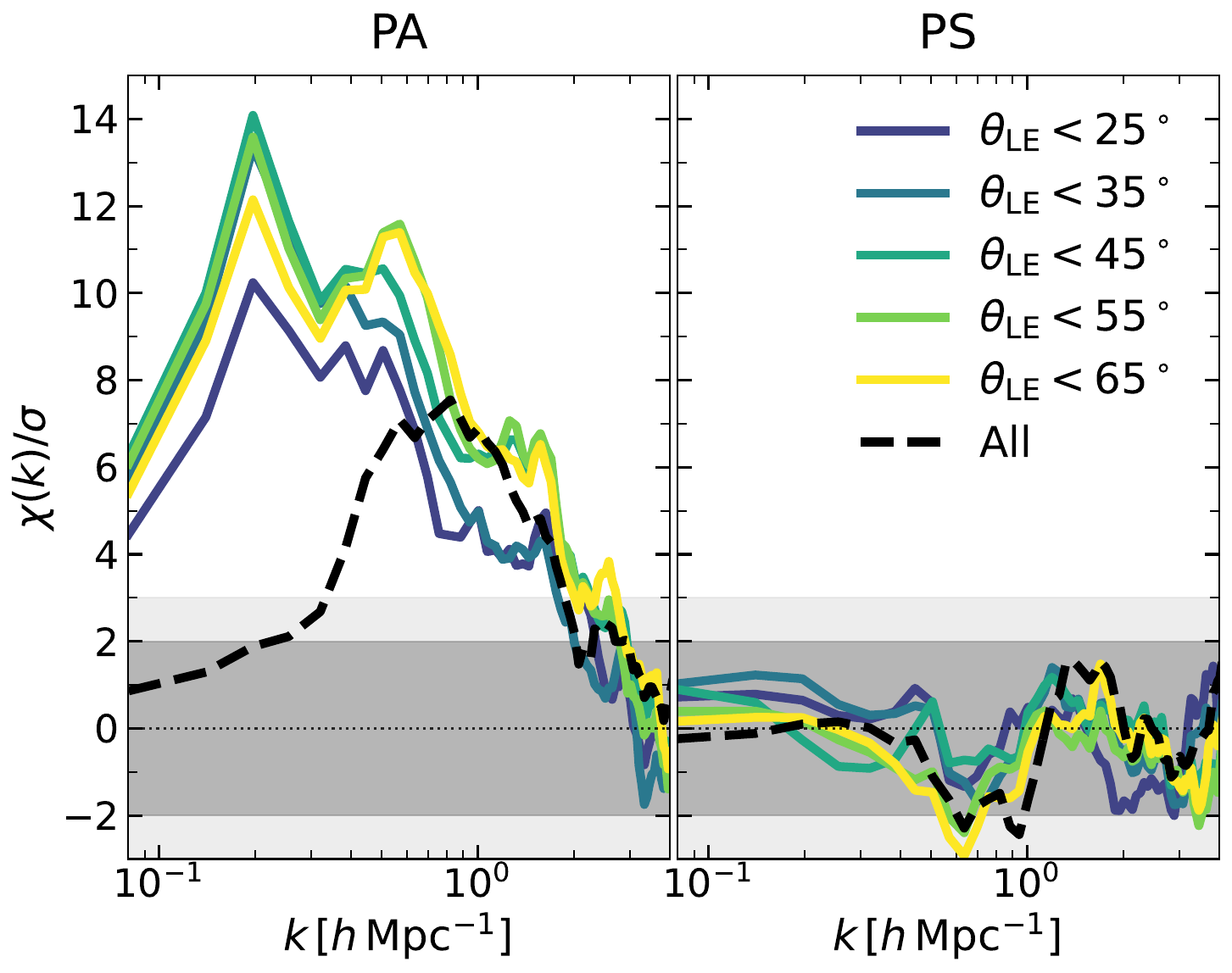}
\caption{Same as Figure~\ref{fig:norm LE results}, but using the full size of each subsample without random sampling for size matching. In the parity-asymmetric simulations, the subsample with $\theta_{\rm LE}<45^\circ$ exhibits the highest detection significance.}
\label{fig:PS_LE_full}
\end{figure}

In Figure~\ref{fig:PS_LE_full}, we now show the result obtained using the full halo subsample of each maximum allowed angle, without tallying the number of halos.
The helical asymmetry signal in the parity-violating universe is maximized in the case for $\theta_{\rm LE} < 45^\circ$ ($\sim 36\%$ of the population), thanks to its larger sample size, exceeding that of the more refined $\theta_{\rm LE} < 25^\circ$ sample.
However, beyond the spin alignment threshold of $45^\circ$, smaller mass halos with weaker LE alignment are more likely to be included, degrading the signal.
We confirmed that this trend is consistently observed across all mass thresholds, as shown in Figure~\ref{fig:LEallmass}.
Based on these findings, we conclude that halos with $\theta_{\rm LE} < 45^\circ$ represent an optimal target halo sample for our subsequent analysis.

\begin{figure*}
\centering
\includegraphics[width=0.95\textwidth]{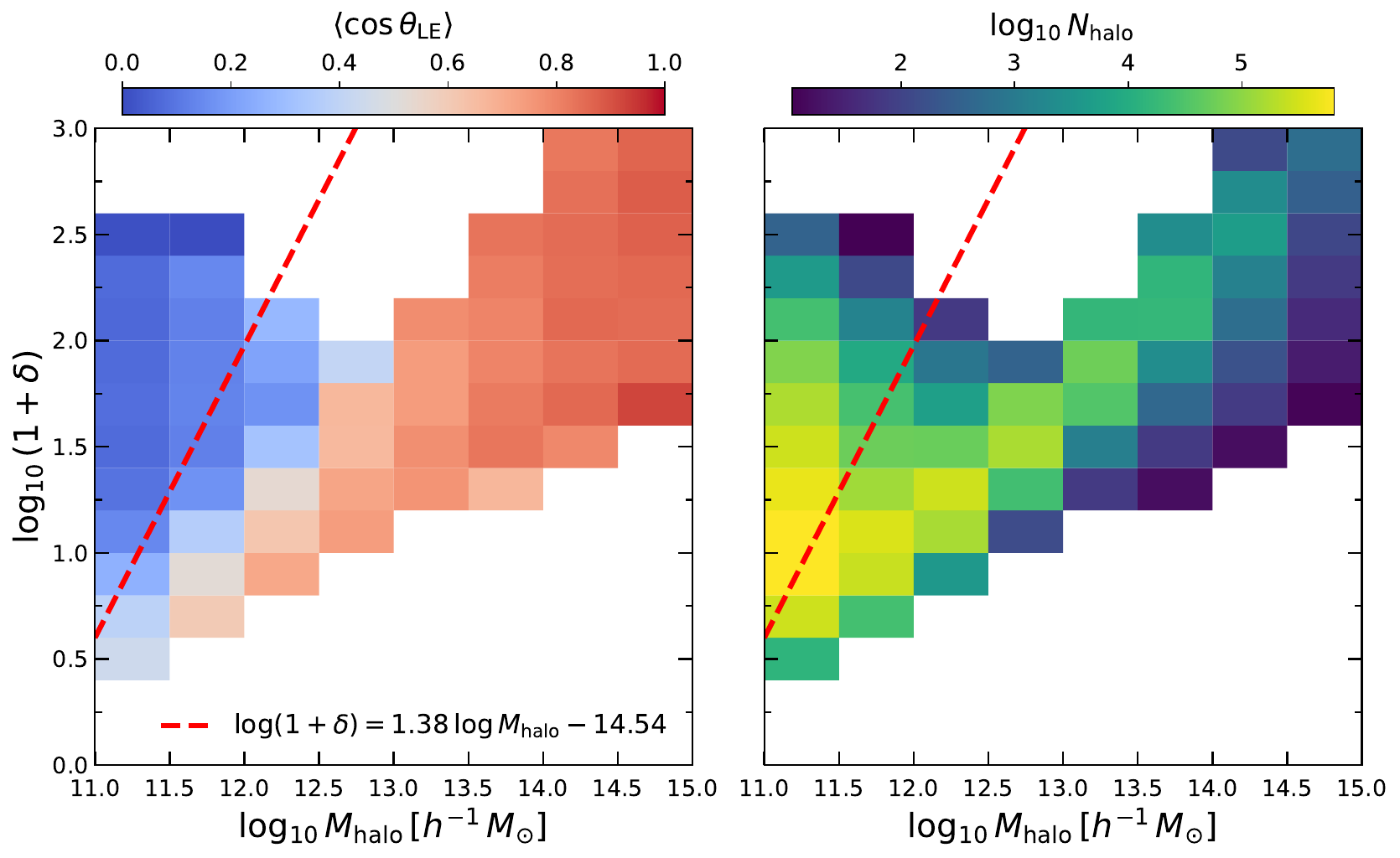}
\caption{Mean LE alignment (left) and the number of halos (right) in the plane of halo mass and local density ($1+\delta$). The local density is estimated using a density field smoothed with a Gaussian kernel of $0.5\,h^{-1} {\rm Mpc}$ scale. The dashed lines indicate the optimal halo selection criterion, with the functional form provided in the legend.}
\label{fig:LE_alignment_in_2D_histogram}
\end{figure*}

\subsection{A Practical Selection Criterion for Strong LE Alignment in the Halo Mass-Local Density Plane}\label{LE alignment in the Mass-Density Plane}
Since LE alignment cannot be directly determined from observations, we instead seek observable halo properties that correlate with it and can be used for practical halo selection.
In large galaxy surveys, quantities that can be derived from observations include halo mass, inferred from galaxy luminosity, and the local density contrast, estimated from the galaxy distribution.
Therefore, we investigate the mean LE alignment of halos in the two-dimensional plane of these two quantities.

Figure~\ref{fig:LE_alignment_in_2D_histogram} shows the mean LE alignment of halos (left panel) and the number of halos (right panel) in each halo mass--local density bin. The local density is calculated using the cloud-in-cell method, smoothed on a scale of $0.5\,h^{-1}\,{\rm Mpc}$, which corresponds to the Gaussian radius associated with the adopted mass threshold of $10^{11}\,h^{-1}\,M_{\odot}$.\footnote{The conclusion of this section is robust against variations in the smoothing scale, provided that the smoothing length is not chosen so large that it homogenizes the local density field too much.} We find that halos exhibit stronger LE spin alignment with increasing halo mass and decreasing local density, while the number of those halos becomes smaller. Such environmental and halo mass dependence of LE alignment is consistent with the observation that halos in underdense regions tend to experience fewer tidal interactions and mergers \citep{Fakhouri_Ma_2009}, better keeping the initial conditions, and that more massive halos better preserve their initial spin orientations \citep{Yu_2020, Wu_2021}, likely because their evolution remains relatively more in the quasi-linear regime.

\begin{figure}
\includegraphics[width=\linewidth]{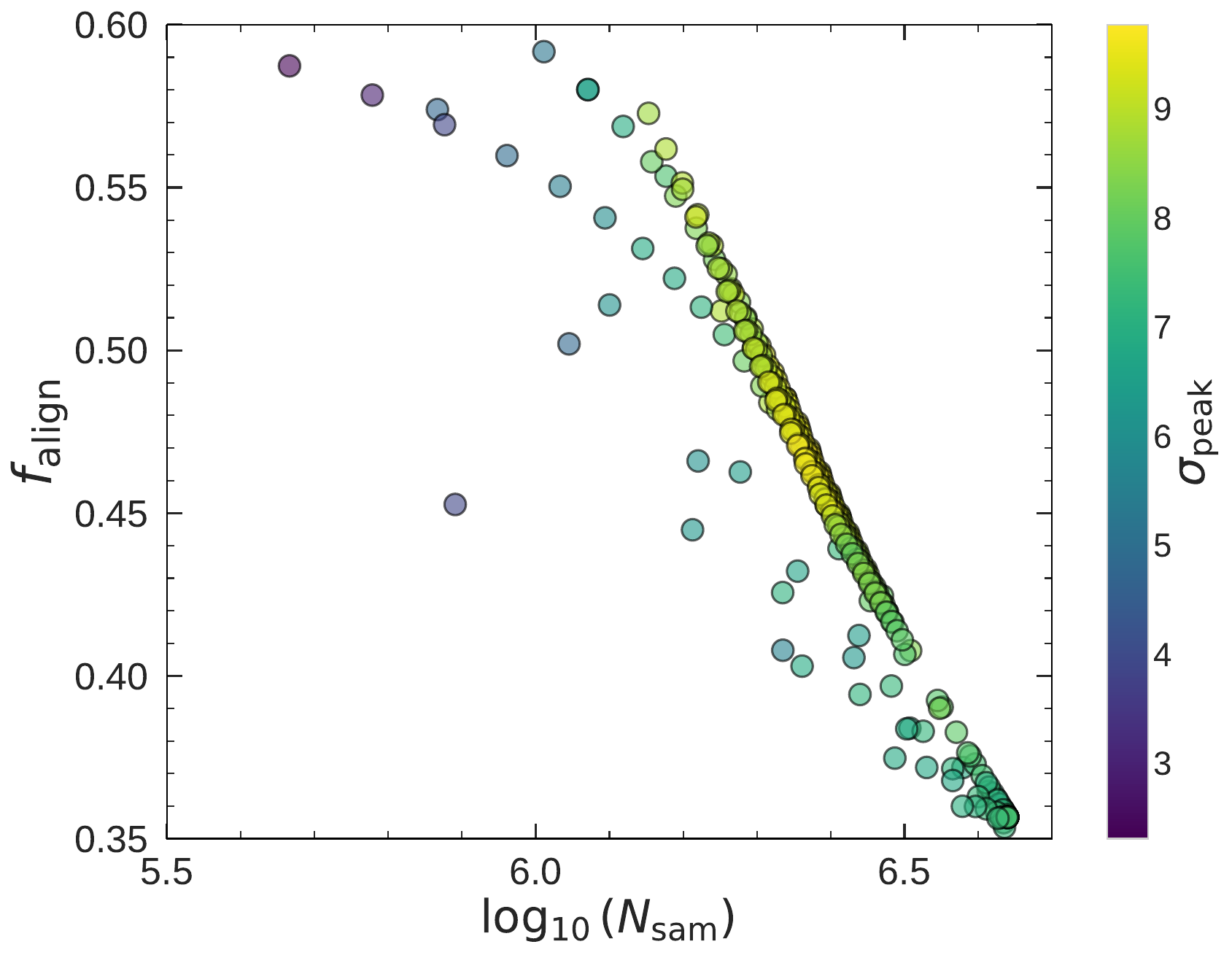}
\caption{The number of halos ($N_{\rm sam}$) and the fraction of halos with strong LE alignment ($f_{\rm align}$; $\theta_{\rm LE}<45^\circ$) of various subsamples explored in the systematic search for an optimal halo selection (see text), color-coded by the peak detection significance of the helical asymmetry power spectrum of each subsample. The optimal halo selection criterion, as depicted in Figure~\ref{fig:LE_alignment_in_2D_histogram}, is derived by balancing the trade-off between $N_{\rm sam}$ and $f_{\rm align}$.}
\label{fig:peak_and_number}
\end{figure}

We then define a dividing line on the halo mass--local density plane to construct a halo sample with a large sample size and tight LE alignment. We search for the optimal line by varying its slope and intercept, changing the total sample size, $N_{\rm sam}$, and the fraction of halos with strong LE alignment, $f_{\rm align}$. 
In Figure~\ref{fig:peak_and_number}, we compute the detection significance as a function of $N_{\rm sam}$ and $f_{\rm align}$ for each halo sample that lies to the right of the dividing line (in Figure~\ref{fig:LE_alignment_in_2D_histogram}).
We note that $N_{\rm sam}$ and $f_{\rm align}$ cannot be varied independently without changing the other; therefore, we can only explore the limited regions in the halo mass--local density plane.
Then we determine the optimal line as the one that yields the highest detection significance. As mentioned in the previous section,  when lowering the halo mass threshold, we find an anti-correlation between $N_{\rm sam}$ and $f_{\rm align}$, i.e., increasing $N_{\rm sam}$ toward decreasing $f_{\rm align}$. Consequently, the detection significance is maximized when the effects of the two factors balance.

\begin{figure}
\centering
\includegraphics[width=0.8\linewidth]{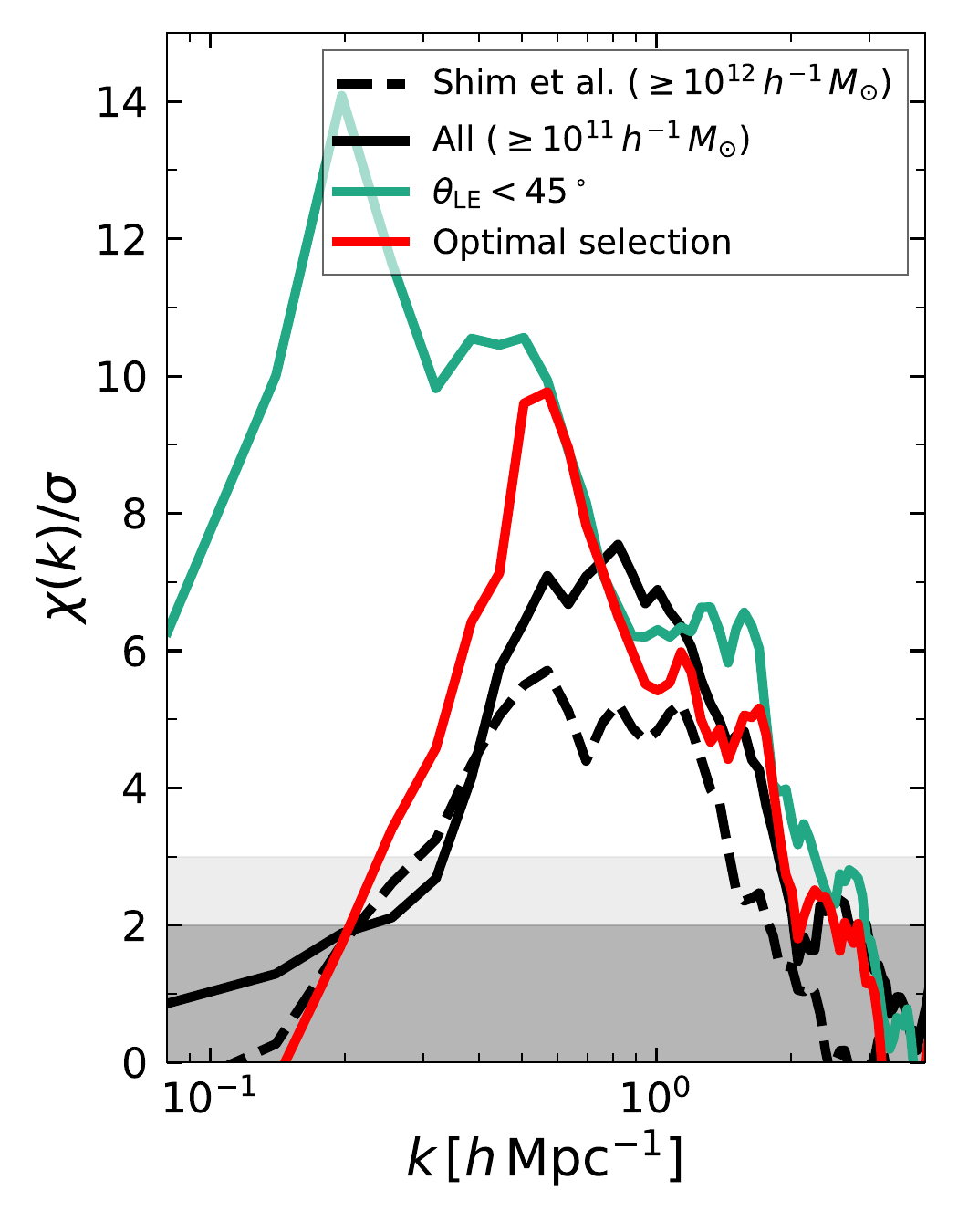}
\caption{Comparison of the detection significances of the helical asymmetry power spectra in PA simulations. Results are shown for the previous study ($M_{\rm halo}\ge10^{12}\,h^{-1}M_\odot$, black dashed) and the current $10^{11}\,h^{-1}M_\odot$ mass-threshold sample (black solid). In addition to the $10^{11}\,h^{-1}M_\odot$ mass cut, we show the results for samples further constrained by strong LE alignment ($\theta_{\rm LE}$ $< 45^\circ$; green) and our optimized criteria (red).}
\label{fig:fine+all+45 power spectra}
\end{figure}

In Figure~\ref{fig:fine+all+45 power spectra}, we compare the level of detection from various halo samples considered in our analysis, including the case presented in \citet{Shim_2025}. Increasing the sample size by lowering the mass threshold to $M\ge10^{11}h^{-1}M_{\odot}$ yields higher detection significance than the case for $M\ge10^{12}h^{-1}M_{\odot}$. For the optimal halo sample we constructed, i.e., a refined expanded halo sample additionally considering high LE alignment criteria, we observe another $29\%$ improvement from the sample with $M\ge10^{11}h^{-1}M_{\odot}$, corresponding to $71\%$ enhancement in comparison to \citet{Shim_2025}.

\section{Observational Detectability of the Parity-Violating Signal} \label{sec: Observation of the PV Signal}

In this section, we forecast how significant the detected signal could be when applying our approach to the current/planned galaxy surveys. We assume galaxy samples from two galaxy surveys: the Dark Energy Spectroscopic Instrument Bright Galaxy Survey (DESI BGS) Bright \citep{Hahn_2023} as the state-of-the-art survey, and the Korean Large Spectroscopic Telescope Hubble Depth Survey (KLST HDS) as the deeper next-generation spectroscopic survey. The final release of the DESI BGS is planned to contain $\sim10^7$ galaxies brighter than $19.5$ magnitude in $r$-band, and the KLST HDS will include dimmer galaxies down to $20.0$ magnitude in $i$-band, expected to provide $\sim 5\times10^7$ galaxies.

To obtain the number and mass distributions of halos that correspond to the observed galaxies in both surveys, we utilize the Santa Cruz Semi-Analytic Model ultra-wide lightcone data \citep[hereafter SC-SAM ultra-wide;][]{Yung_2022b}. The SC-SAM data have been tested against observations over a wide redshift range, and shown to reproduce key observed galaxy properties, including stellar mass functions, star formation rates, UV luminosity functions, and empirical galaxy--halo scaling relations \citep{Somerville_2015, Yung_2019a, Yung_2019b, Gabrielpillai_2022}.

\begin{figure*}
\includegraphics[width=\linewidth]{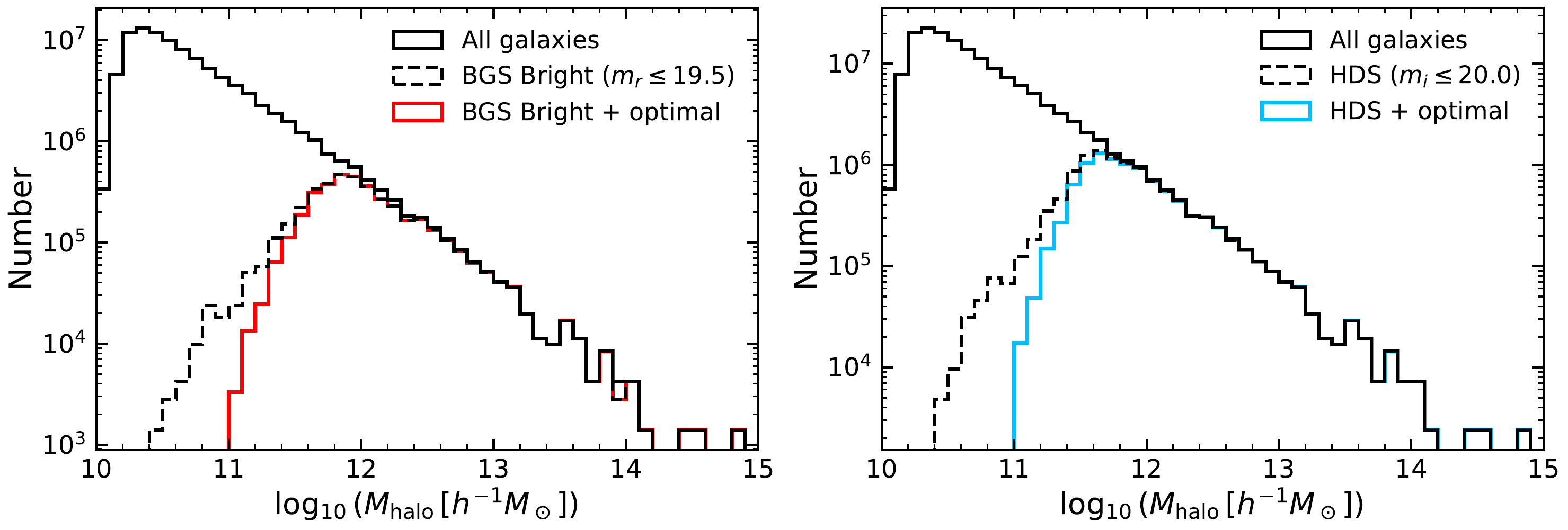}
\caption{Halo mass distributions for galaxies at $z\le0.2$ in the SC-SAM ultra-wide mocks. The number of halos are scaled to match the footprint of each survey. The black solid lines represent all galaxies in the full mock samples, while the dashed lines show the observable mock samples for BGS Bright (left) and HDS (right). The colored histograms (red and blue) indicate the expected number of galaxies satisfying the optimal selection criteria (see Section~\ref{LE alignment in the Mass-Density Plane}), estimated by applying the fraction of simulation halos selected in each mass bin to the corresponding observable mock catalogs.}
\label{fig:observational feasibility}
\end{figure*}

In Figure~\ref{fig:observational feasibility}, we show the mass distributions of entire halos hosting galaxies in SC-SAM ultra-wide (black solid) and their subsamples as the mocks (black dashed) for the final DESI BGS Bright (left panel) and the KLST HDS (right panel), considering their observational magnitude limits. The number and mass distributions of the observationally accessible halo subsamples are obtained through simple scaling between the footprints of the SC-SAM ultra-wide and galaxy surveys. For our forecast, we adopt a conservative upper redshift limit of $z\le0.2$ for the reliable spin determination from imaging data \citep{Shamir_2020, Shamir_2022, Jia_2023}.
Rather than explicitly imposing our environmental selection on this sample, we apply the fraction of halos satisfying our criteria at each halo mass bin---as derived from the PS/PA simulations---to the observable population (red and blue).
Consequently, we expect to secure a total of $\sim 3.8\,\text{million}$ galaxies satisfying our optimal selection criteria, yielding an expected significance of $11.85\sigma$. Applying the same selection strategy to the HDS mock sample yields a significance of $19.14\sigma$.

Overall, these results suggest that both surveys have the potential to enable a statistically significant detection of the helical asymmetry signal for optimally selected halos. Although our estimates do not explicitly account for galaxy--halo spin misalignment and observational error, the detection significance would remain above $9\sigma$ for DESI BGS and $15\sigma$ for HDS Wide,\footnote{We applied a simple reduction of $18\%$ to the estimated detection significance following \citep{Shim_2025}, which accounts for the galaxy--halo spin misalignment and observational error.} indicating that a robust detection is achievable with both surveys.

\section{Summary and Conclusions} \label{sec:summary}
In this study, we explore a strategy to enhance the detection significance of primordial parity violation using halo spins beyond the level achieved in \citet{Shim_2025}. To this end, we construct a spin field from halos optimally selected according to their mass and local density, while extending the halo mass range beyond that adopted in the previous study, which is limited to halos with $M_{\rm halo} \ge 10^{12}\,h^{-1}\,M_\odot$. Applying a simple mass cut, we find that the highest detection significance is obtained for a threshold of $10^{11.5}\,h^{-1}\,M_\odot$, while a threshold of $10^{11}\,h^{-1}\,M_\odot$ also performs better than the other thresholds (Figure~\ref{fig:masscut}). However, when restricting the sample to halos with strong LE alignment ($\theta_{\rm LE}<45^\circ$), the $10^{11}\,h^{-1}\,M_\odot$ threshold yields the highest significance (Figures~\ref{fig:PS_LE_full} and \ref{fig:LEallmass}).
The halos optimally selected in the halo mass--local density plane yield $29\%$ and $71\%$ improvement over the simple mass-cut case and the previous study, respectively (Figure~\ref{fig:fine+all+45 power spectra}).

We further assessed the observational feasibility of detecting this primordial parity-violating signal using galaxy spins. Utilizing the SC-SAM ultra-wide, we evaluate the accessible sample sizes and expected detection significances under the specific constraints of the final DESI BGS Bright and the planned HDS samples. By applying our optimal halo selection, both surveys are expected to provide sufficiently large samples to enable a statistically significant detection, even when accounting for signal degradation due to galaxy-halo spin misalignment and observational uncertainties. This demonstrates that a highly robust detection of the primordial parity-violating signal is achievable with ongoing and upcoming spectroscopic surveys.

While the current framework relies solely on halo mass and local density, a natural next step is to extend this parameter space by incorporating the information on halo tidal environments. Since tidal interactions play a key role in shaping the evolution and reorientation of halo spins \citep{Wu_2021, Moon_Lee_2024}, environmental indicators derived from the local tidal field may provide additional power for identifying halos that preserve primordial spin information. This may further enhance the detectability of primordial parity-violating signals, an avenue that we plan to explore in future work.

\section*{Acknowledgements}
This work was supported by the National Research Foundation of Korea (NRF) grant funded by the Korean government (MSIT) (No. RS-2026-25478560). We thank ASIAA for providing computing resources (High Performance Computing Systems). This research was supported by the Global - Learning \& Academic research Institution for Master's $\cdot$ PhD students, and Postdocs (G-LAMP) Program of the National Research Foundation of Korea (NRF) grant funded by the Ministry of Education (No. RS-2025-25442707).

\section*{Data availability}
The data underlying this article will be shared on reasonable request to the corresponding author.

\bibliography{ref}{}
\bibliographystyle{aasjournal}

\appendix
\counterwithin{figure}{section}
\renewcommand{\thefigure}{\thesection\arabic{figure}}

\FloatBarrier
\section{Alignment effect at different mass threshold}\label{appen}
\begin{figure}
\includegraphics[width=\textwidth, height=0.7\textheight, keepaspectratio]{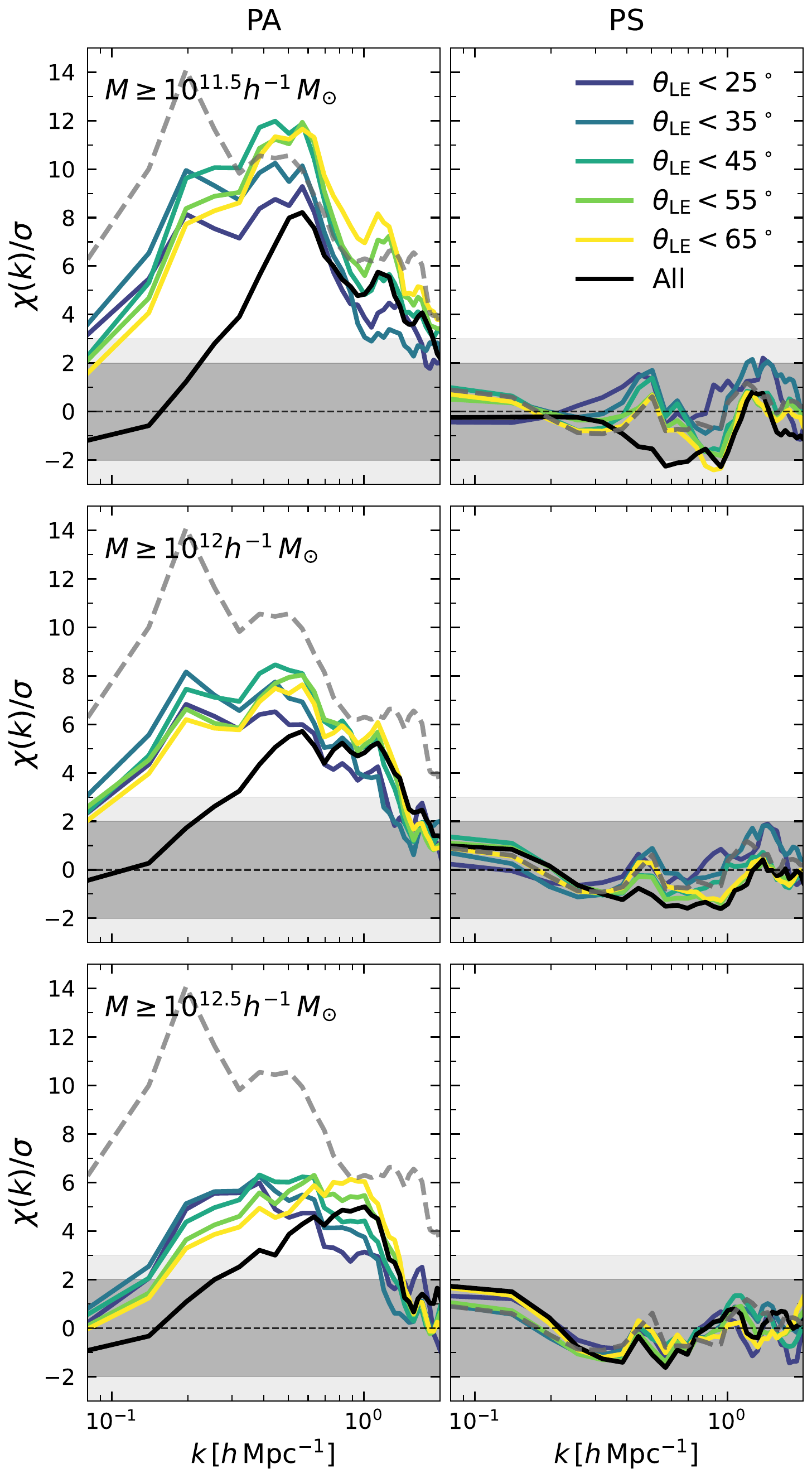}
\caption{Same analysis as in Figure~\ref{fig:PS_LE_full}, but for different mass thresholds. The dashed lines represent the case of the sample with $M_{\rm halo} \ge 10^{11}\,h^{-1}$ and $\theta_{\rm LE}<45^\circ$, as shown in Figure~\ref{fig:PS_LE_full}.}
\label{fig:LEallmass}
\end{figure}

Figure~\ref{fig:LEallmass} extends the LE alignment analysis of Section~\ref{sec: Signal enhancement through early-late halo spin alignment selection} to higher halo mass thresholds. In all cases, restricting the sample to halos with strong LE alignment ($\theta_{\rm LE} < 45^\circ$) consistently yields the highest detection significance. Furthermore, the $10^{11}\,h^{-1}\,M_\odot$ sample with $\theta_{\rm LE} < 45^\circ$ (gray dashed line) outperforms the entire results at all higher thresholds, confirming that this mass cut combined with the LE alignment criterion represents the optimal selection for maximizing the detection significance of the primordial PV signal.

\end{document}